# Quantum Computing and Phase Transitions in Combinatorial Search

Tad Hogg

Dynamics of Computation Group
Xerox Palo Alto Research Center
3333 Coyote Hill Road
Palo Alto, CA 94304
hogg@parc.xerox.com

## Abstract

We introduce an algorithm for combinatorial search on quantum computers that is capable of significantly concentrating amplitude into solutions for some NP search problems, on average. This is done by exploiting the same aspects of problem structure as used by classical backtrack methods to avoid unproductive search choices. This quantum algorithm is much more likely to find solutions than the simple direct use of quantum parallelism. Furthermore, empirical evaluation on small problems shows this quantum algorithm displays the same phase transition behavior, and at the same location, as seen in many previously studied classical search methods. Specifically, difficult problem instances are concentrated near the abrupt change from underconstrained to overconstrained problems.

August 16, 1995

# 1. Introduction

Computation is ultimately a physical process [31]. That is, in practice the range of physically realizable devices determines what is computable and the resources, such as computer time, required to solve a given problem. Computing machines can exploit a variety of physical processes and structures to provide distinct trade-offs in resource requirements. An example is the development of parallel computers with their trade-off of overall computation time against the number of processors employed. Effective use of this trade-off can require algorithms that would be very inefficient if implemented serially.

Another example is given by hypothetical quantum computers [11]. They offer the potential of exploiting quantum parallelism to trade computation time against the use of coherent interference among very many different computational paths. However, restrictions on physically realizable operations make this trade-off difficult to exploit for search problems, resulting in algorithms essentially equivalent to the inefficient method of generate-and-test. Fortunately, recent work on factoring [42] shows that better algorithms are possible. Here we continue this line of work by introducing a new quantum algorithm, motivated by classical backtrack search methods, for a class of particularly difficult combinatorial search problems. Interestingly, while this algorithm represents a substantial improvement for quantum computers, it is particularly inefficient as a classical search method, both in memory and time requirements.

When evaluating algorithms, computational complexity theory usually focuses on the scaling behavior in the worst case. Of particular theoretical concern is whether the search cost grows exponentially or polynomially. However, in many practical situations, typical or average behavior is of more interest. This is especially true because many instances of search problems are much easier to solve than is suggested by worst case analyses. In fact, recent studies have revealed a more detailed structure of the class of search problems. Specifically, for a wide variety of classical search methods, the hard instances are not only rare but also concentrated near abrupt transitions in problem behavior analogous to physical phase transitions [25]. In order to exhibit this concentration of hard instances a search algorithm must exploit the problem constraints to prune unproductive search choices. Unfortunately, this is not easy to do within the range of allowable quantum computational operations. It is thus of interest to see if these results generalize to quantum search methods as well.

In this paper, the new algorithm is evaluated empirically to determine its average behavior. The algorithm is also shown to exhibit the phase transition, indicating it is indeed managing to, in effect, prune unproductive search. This leaves for future work the analysis of its worst case performance.

This paper is organized as follows. First we discuss combinatorial search problems and the phase transitions where hard problem instances are concentrated. Second, after a brief summary of quantum computing, the new quantum search algorithm is motivated and described. In fact, there are a number of natural variants of the general algorithm. Two of these are evaluated empirically on a range of problems to exhibit the generality of the phase transition and their performance. Finally, some important caveats for the implementation of quantum computers and open issues are presented.

## 2. Combinatorial Search

Combinatorial search is among the hardest of common computational problems: the solution time can grow exponentially with the size of the problem [20]. Examples arise in scheduling, planning, circuit layout and machine vision, to name a few areas. Many of these examples can be viewed as constraint satisfaction problems [34]. Here we are given a set of *n* variables each of which can be assigned *b* possible values. The problem is to find an assignment for each variable that together satisfy some specified constraints. For instance, consider the small scheduling problem of selecting one of two periods in which to teach each of two classes that are taught by the same person. We can regard each class as a variable and its time slot as its value, i.e., here $n = b = 2$. The constraints are that the two classes are not assigned to be at the same time. Note that having a nontrivial search requires at least two possible values to assign to each variable, so we restrict attention to cases where $b \geq 2$.

Fundamentally, the combinatorial search problem consists of finding those combinations of a discrete set of items that satisfy specified requirements. The number of possible combinations to consider grows very rapidly (e.g., exponentially or factorially) with the number of items, leading to potentially lengthy solution times and severely limiting the feasible size of such problems. For example, the number of possible assignments in a constraint problem is $b^n$, which grows exponentially with the problem size (given by the number of variables *n*).

Because of the exponentially large number of possibilities it appears the time required to solve such problems must grow exponentially, in the worst case. However for many such problems it is easy to verify a solution is in fact correct. These problems form the well-studied class of NP problems: informally we say they are hard to solve but easy to check. One well-studied instance is graph coloring, where the variables represent nodes in a graph, the values are colors for the nodes and the constraints are that each pair of nodes linked by an edge in the graph must have different colors. Another example is propositional satisfiability, where the variables take on logical values of true or false, and the assignment must satisfy a specified propositional formula involving the variables. Both these examples are instances of particularly difficult NP problems known as the class of NP-complete search problems [20].

### 2.1. Phase Transitions

Much of the theoretical work on NP search problems examines their worst case behavior. Although these search problems can be very hard, in the worst case, there is a great deal of individual variation in these problems and among different search methods. Recently there have been a number of studies of the structure of the class of NP search problems focusing on regularities of the typical behavior [7, 36, 48, 25, 23]. This work has identified a number of common behaviors. Specifically, for large problems, a few parameters characterizing their structure determine the relative difficulty for a wide variety of common search methods, on average. Moreover, changes in these parameters give rise to transitions, becoming more abrupt for larger problems, that are analogous to phase transitions in physical systems. In this case, the transition is from underconstrained to overconstrained problems, with the hardest cases concentrated near the transition region of critically constrained problems. One powerful result of this work is that this concentration of hard cases occurs at the same parameter values for a wide range of search methods. That is, this behavior is a property of the problems rather than of the details of the search algorithm.

This can be understood by viewing a search as making a series of choices until a solution is found. The overall search will usually be relatively easy (i.e., require few steps) if either there are many choices leading to solutions or else choices that do not lead to solutions can be recognized quickly as such, so that unproductive search is avoided. Whether this condition holds is in turn determined by how tightly constrained the problem is. When there are few constraints almost all choices are good ones, leading quickly to a solution. With many constraints, on the other hand, there are few good choices but the bad ones can be recognized very quickly as violating some constraints so that not much time is wasted considering them. In between these two cases are the hard problems: enough constraints so good choices are rare but few enough that bad choices are usually recognized only with a lot of additional search.

A more detailed analysis suggests a series of transitions [26]. With very few constraints, the average search cost scales polynomially. As more constraints are added, there is a transition to exponential scaling. The rate of growth of this exponential increases until the transition region described above is reached. Beyond that point, with its concentration of hard problems, the growth rate decreases. Eventually, for very highly constrained problems, the search cost again grows only polynomially with size.

## 2.2. The Combinatorial Search Space

A general view of the combinatorial search problem is that it consists of a number $N$ of items and a requirement to find a solution, i.e., a set of $L < N$ items that satisfies specified requirements. These requirements in turn can be described as a collection of *nogoods*, i.e., sets of items whose combination is not allowed. For the situation studied here, the nogoods directly specified by the problem requirements will be small sets of items, e.g., of size two or three. On the other hand, the number of items and the size of the solutions will grow with the problem size. A key point that makes this set representation conceptually useful is that if a set is nogood, so are all of its supersets.

For the case of constraint satisfaction, the items are just the possible assignments. Thus there are $N = nb$ items[1]. A solution consists of an assignment to each variable that satisfies whatever constraints are given in the problem. Thus a solution consists of a set of $L = n$ items. In terms of the general framework for combinatorial search these constraint satisfaction problems will also contain a number of problem-independent *necessary* nogoods, namely those corresponding to giving the same variable two different values. There are $n \binom{b}{2}$ such necessary nogoods. For a nontrivial search we must have $b \geq 2$, so we restrict our attention to the case where $L \leq N/2$. This requirement is important in allowing the construction of the quantum search method described below.

---

[1] The lattice of sets can also represent problems where each variable can have a different number of assigned values.

In this context we define a *good* to be a set of items that is consistent with all the constraints of the problem, while a *nogood* is an inconsistent set. We also say a set is *complete* if it has $L$ items, while smaller sets are *partial* or *incomplete*. Thus a solution corresponds to a complete good set. In addition, a *partial solution* is an incomplete good set. These sets, grouped by size and with each set linked to its immediate supersets and subsets, form a lattice structure. This structure for $N = 4$ is shown in Fig. 1. We say that the

$$N_i = \binom{N}{i} \tag{1}$$

sets of size $i$ are at level $i$ in the lattice.

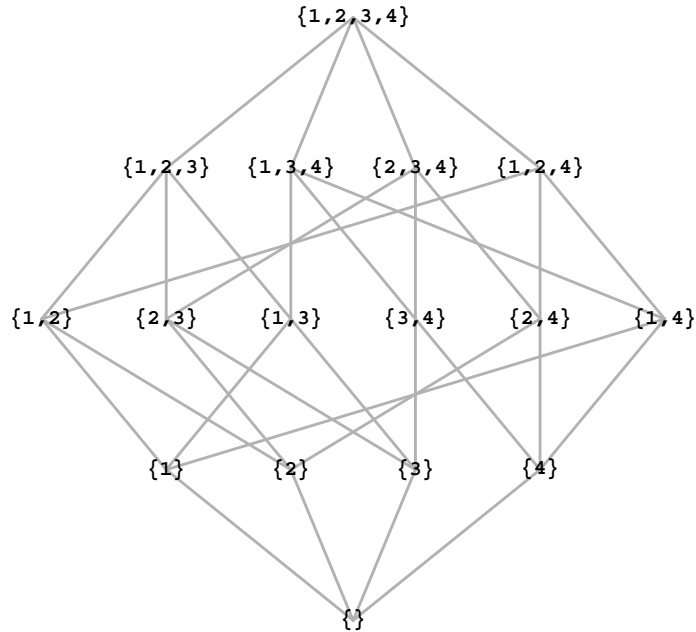

**Fig. 1.** Structure of the set lattice for a problem with four items. The subsets of $\{1, 2, 3, 4\}$ are grouped into levels by size and lines drawn between each set and its immediate supersets and subsets. The bottom of the lattice, level 0, represents the single set of size zero, the four points at level 1 represent the four singleton subsets, etc.

As an example, consider a problem with $N = 4$ and $L = 2$, and suppose the constraints eliminate items 1 and 3. Then we have the sets $\{\}$, $\{2\}$, and $\{4\}$ as partial goods, while $\{1\}$ and $\{3\}$ are partial nogoods. Among the 6 complete sets, only $\{2,4\}$ is good as the others are supersets of $\{1\}$ or $\{3\}$ and hence nogood.

In fact, we generally do not expect to see search problems whose constraints explicitly involve nogoods of size 0 or 1. This is because a nogood of size 0, i.e., the empty set, immediately makes all sets inconsistent, giving a trivially insoluble problem. Similarly, a nogood of size 1 just eliminates the corresponding item from consideration hence easily transforming the problem into a smaller one with that item removed from consideration. Thus we restrict attention to cases where the problem requirements involve at least two items at a time. On the other hand, if the problem requirements only involve large sets of items, nogoods won't appear until relatively close to the solution level in the lattice, resulting is less opportunity to guide the search based on evaluating sets of intermediate size. In this paper we focus on problems whose requirements explicitly involve only a few items at a time, but at least two. This gives a number of small nogoods, i.e., near the bottom of the lattice. Examples of such problems include binary constraint satisfaction, graph coloring and propositional satisfiability mentioned above.

An example is given by a simple constraint satisfaction problem consisting of $n = 2$ variables ($v_1$ and $v_2$) each of which can take on one of $b = 2$ values (1 or 2) and the single constraint that the two variables take on distinct values, i.e. $v_1 \neq v_2$. Hence there are $N = nb = 4$ assignments: $v_1 = 1, \ v_1 = 2, \ v_2 = 1, \ v_2 = 2$ which we denote as items $1, 2, 3, 4$ respectively. The corresponding lattice is given in Fig. 1. What are the nogoods for this problem? First there are those due to the explicit constraint that the two variables have distinct values: $\{v_1 = 1, \ v_2 = 1\}$ and $\{v_1 = 2, \ v_2 = 2\}$ or $\{1, 3\}$ and $\{2, 4\}$. In addition, there are necessary nogoods implied by the requirement that a variable takes on a unique value so that any set giving multiple assignments to the same variable is necessarily nogood, namely $\{v_1 = 1, \ v_1 = 2\}$ and $\{v_2 = 1, \ v_2 = 2\}$ or $\{1, 2\}$ and $\{3, 4\}$. Referring to Fig. 1, we see that these four nogoods force all sets of size 3 and 4 to be nogood too. However, sets of size zero and one are goods as are the remaining two sets of size two: $\{2, 3\}$ and $\{1, 4\}$ corresponding to $\{v_1 = 2, \ v_2 = 1\}$ and $\{v_1 = 1, \ v_2 = 2\}$ which are the solutions to this problem.

Various search methods correspond to different strategies for examining the sets in this lattice. For instance, methods such as simulated annealing [30], heuristic repair [35] and GSAT [41] move among complete sets, attempting to find a solution by a series of small changes to the sets. Generally these search techniques continue indefinitely if the problem has no solution and thus they can never show that a problem is insoluble. Such methods are called *incomplete*. In these methods, the search is repeated over and over again, from different initial conditions or making different random choices, until either a solution is found or some specified limit on the number of trials is reached. In the latter case, one cannot distinguish a problem with no solution at all from just a series of unlucky choices for a soluble problem. Other search techniques attempt to build solutions starting from smaller sets, often by a process of extending a consistent set until either a solution is found or no further consistent extensions are possible. In the latter case the search backtracks to a previous decision point and tries another possible extension until no further choices remain. By recording the pending choices at each decision point, these backtrack methods can determine a problem is insoluble, i.e., they are *complete* or *systematic* search methods.

This description highlights two distinct aspects of the search procedure: a general method for moving among sets, independent of any particular problem, and a testing procedure that checks sets for consistency with the particular problem's requirements.

Often, heuristics are used to make the search decisions depend on the problem structure hoping to identify changes most likely to lead to a solution. However, conceptually these aspects can be separated, as in the case of the quantum search algorithm presented below.

For constraint satisfaction problems, a simple alternative representation to this lattice structure is to use partial assignments, i.e., sets of assignments guaranteed to give at most one value to each variable. At first sight this might seem better in that it removes from consideration many sets guaranteed to be nogood (i.e., those with multiple assignments to some variable) and hence increases the proportion of complete sets that are solutions. However, in this case the number of sets as a function of level in the lattice would decrease before reaching the solution level, precluding the simple form of a unitary mapping described below for the quantum search algorithm. Another representation that avoids this problem is to consider assignments in only a single arbitrary order. This version of the set lattice has been previously used in theoretical analyses of search [48]. This may be useful to explore further for the quantum search, but is unlikely to be as effective. This is because some sets will become nogood only at the last few steps in a fixed ordering, resulting is less opportunity to use intermediate size nogoods to focus on solutions.

## 3. Quantum Search Methods

This section briefly describes the capabilities of ideal quantum computers, why some straightforward attempts to exploit these capabilities for search are not particularly effective, then motivates and describes a new search algorithm

### 3.1. An Overview of Quantum Computers

The basic distinguishing feature of a quantum computer [3–5, 11, 12, 16, 17, 27, 29, 33, 42, 46] is its ability to operate simultaneously on a collection of classical states, thus potentially performing very many operations in the time a classical computer would do just one. Alternatively, this can be viewed as a large parallel computer but requiring no more hardware than that needed for a single processor. More specifically, suppose $s_1, \ldots, s_N$ are possible states for a classical computer, e.g., these could represent the possible states of a register consisting of *n* bits with $N = 2^n$. The corresponding state of the quantum computer is described by a linear superposition of these classical states $|s\rangle$, each with an associated complex number called its amplitude, i.e., $|s\rangle = \sum \psi_i |s_i\rangle$. Here we use the standard ket notation from quantum mechanics [13, section 6] to denote various states, and to distinguish these states from the amplitudes[2]. These superpositions can also be viewed as vectors in a space whose basis is the individual classical states $|s_i\rangle$ and $\psi_i$ is the component of the vector along the i[th] basis element of the space. Such a vector can also be specified by its components as $(\psi_1, \ldots, \psi_N)$.

Consider some reversible classical computation on these states, e.g., $f(s_i)$. When applied to a superposition of states, the result is $|f(s)\rangle = \sum \psi_i |f(s_i)\rangle$, that is, operations behave linearly with respect to superpositions, a key principle of quantum mechanics. In addition to this use of computations on individual states, there are two other processes. First, it is possible to create new superpositions by operating on a given one with a unitary matrix[3], i.e., $U|s\rangle = \sum \psi_i U|s_i\rangle$. Second, a measurement can be made on the system to select out one of the states in the superposition. When a measurement is made on the quantum computer, e.g., to determine the result of the computation represented by a particular configuration of the bits in a register, one of the possible classical states is obtained. Specifically, the state $s_i$ is obtained with probability $|\psi_i|^2$, thus giving the physical interpretation of the amplitudes. The requirement for unitarity arises from the physical requirement that the probabilities of any superposition of states must always sum to one, i.e., $\sum_i |\psi_i|^2 = 1$ (in particular, this means the mapping must be reversible). A classical computer can then be viewed as a special case of a quantum computer in which at all times exactly one of the $\psi_i$ is equal to one and the rest are zero.

---
[2] The ket notation is conceptually similar to the use of boldface to denote vectors and distinguish them from scalars.

A simple example of these ideas is given by considering a single bit. In this case there are two possible classical states $|0\rangle$ and $|1\rangle$ corresponding to the values 0 and 1, respectively, for the bit. This defines a two dimensional vector space of superpositions for a quantum bit. A simple computation on such a bit is the logical NOT operation, i.e., $\text{NOT}(|0\rangle) = |1\rangle$ and $\text{NOT}(|1\rangle) = |0\rangle$. This operator simply exchanges the vector's components:

$$\text{NOT}\begin{pmatrix} \psi_0 \\ \psi_1 \end{pmatrix} \equiv \text{NOT}(\psi_0|0\rangle + \psi_1|1\rangle) = \psi_0|1\rangle + \psi_1|0\rangle \equiv \begin{pmatrix} \psi_1 \\ \psi_0 \end{pmatrix} \quad (2)$$

This operation can also be represented as multiplication by the permutation matrix $\begin{pmatrix} 0 & 1 \\ 1 & 0 \end{pmatrix}$. Another operator is given by the rotation matrix

$$U(\theta) = \begin{pmatrix} \cos\theta & -\sin\theta \\ \sin\theta & \cos\theta \end{pmatrix} \quad (3)$$

This can be used to create superpositions from single states, e.g.,

$$U\left(\frac{\pi}{4}\right)\begin{pmatrix} 1 \\ 0 \end{pmatrix} \equiv U\left(\frac{\pi}{4}\right)|0\rangle = \frac{1}{\sqrt{2}}(|0\rangle + |1\rangle) \equiv \frac{1}{\sqrt{2}}\begin{pmatrix} 1 \\ 1 \end{pmatrix} \quad (4)$$

This illustrates how a simple initial state can be converted into superpositions. Repeating this operation on *n* bits, will give a superposition of all $2^n$ possible values for those bits, requiring just *n* steps. This ability to mix states instead of just permuting them is important in allowing amplitude to be concentrated into desired states. Although only the magnitude of the amplitude matters in determining the probability of measurement, the phase is important for this mixing process as it can result in constructive or destructive interference.

A quantum computation consists of preparing an initial superposition of states, operating on those states with a series of unitary matrices, and then making a measurement to obtain a definite final answer. The amplitudes determine the probability that this final measurement produces a desired result. Using this as a search method, we obtain each final state with some probability, and some of these states will be solutions. Thus this is a probabilistic computation in which at each trial there is some probability to get a solution, but no guarantee. This means the search method is incomplete: it can find a solution if one exists but can never guarantee a solution doesn't exist.

---

[3] A complex matrix $U$ is said to be unitary when $U^\dagger U = I$, where $U^\dagger$ is the transpose of $U$ with all entries changed to their complex conjugates.

A useful conceptual view of these quantum maps is provided by the path integral approach to quantum mechanics [18]. In this view, the final amplitude of a given state is obtained by summing over all possible paths that produce that state, weighted by suitable amplitudes. In this way, the various possibilities involved in a computation can interfere with each other, either constructively or destructively. This differs from the classical combination of probabilities of different ways to reach the same outcome: the probabilities are simply added, giving no possibility for interference. As a simple example, suppose we have a computation that depends on a single choice. The possible choices can be represented as an input bit with value 0 or 1. Suppose the result of the computation from a choice is also a single value, 1 or -1, representing, say, some consequence of the choice. If one is interested in whether the two results are the same, classically this requires evaluating each choice separately. With a quantum computer we could instead prepare a superposition of the inputs, $\frac{1}{\sqrt{2}}(|0\rangle + |1\rangle)$ using the matrix $U\left(\frac{\pi}{4}\right)$, then do the evaluation to give $\frac{1}{\sqrt{2}}(f_0|0\rangle + f_1|1\rangle)$ where $f_i$ is the evaluation from input $i$, and equals 1 or -1. Finally we combine the states again using $U\left(-\frac{\pi}{4}\right)$ to obtain $\frac{1}{2}((f_0 + f_1)|0\rangle + (f_0 - f_1)|1\rangle)$. Now if both choices give the same value for $f$, this result is $\pm|0\rangle$ so the final measurement process will give 0. Conversely, if the values are different this resulting state is $\pm|1\rangle$ and the measurement gives 1. Thus with the effort required to compute one value classically, we are able to determine definitely whether the two evaluations are the same or different. In this example, for the question of interest we were able to arrange to be in a single state at the end of the computation and hence have no probability for obtaining the wrong answer by the measurement. This result is viewed as summing over the different paths. E.g., the final amplitude for $|0\rangle$, was the sum over the paths $|0\rangle \rightarrow |0\rangle \rightarrow |0\rangle$ and $|0\rangle \rightarrow |1\rangle \rightarrow |0\rangle$. The various formulations of quantum mechanics, involving operators, matrices or sums over paths are equivalent but suggest different intuitions about constructing possible quantum algorithms.

### 3.2. Some Approaches to Search

At first sight quantum computers would seem to be ideal for combinatorial search problems that are in the class NP. In such problems, there is an efficient procedure $f(s)$ that takes a potential solution set $s$ and determines whether $s$ is in fact a solution, but there are exponentially many potential solutions, very few of which are in fact solutions. If $s_1, \ldots, s_N$ are the potential sets to consider, we can quickly form the superposition $\frac{1}{\sqrt{N}}(|s_1\rangle + \ldots + |s_N\rangle)$ and then simultaneously evaluate $f(s)$ for all these states, resulting in a superposition of the sets and their evaluation, i.e., $\frac{1}{\sqrt{N}} \sum |s_i, f(s_i)\rangle$. At this point the quantum computer has, in a sense, evaluated all possible sets and determined which are solutions. Unfortunately, if we make a measurement of the system, we get each set with equal probability $1/N$ and so are very unlikely to observe a solution. This is thus no better than the slow classical search method of random generate-and-test where sets are randomly constructed and tested until a solution is found. Alternatively, we can obtain a solution with high probability by repeating this operation $O(N)$ times, either serially (taking a long time) or with multiple copies of the device (requiring a large amount of hardware or energy if, say, the computation is done by using multiple photons). This shows a trade-off between time and energy (or other physical resources), conjectured to apply more generally to solving these search problems [6], and also seen in the trade-off of time and number of processors in parallel computers.

To be useful for combinatorial search, we can't just evaluate the various sets but instead must arrange for amplitude to be concentrated into the solution sets so as to greatly increase the probability a solution will be observed. Ideally this would be done with a mapping that gives constructive interference of amplitude in solutions and destructive interference in nonsolutions. Designing such maps is made complicated by the fact that they cannot be arbitrary functions. Rather, on physical grounds, they must be linear unitary operators as described above. Beyond this physical restriction, there is an algorithmic or computational requirement: the mapping should be efficiently computable. For example, the map cannot require *a priori* knowledge of the solutions (otherwise there would be no point in using the map to do the search!). More generally the matrix elements should be efficiently computable. This computational requirement is analogous to the restriction on heuristics in standard search methods: to be useful, the heuristic itself must not take a long time to compute. These requirements on the mapping trade off against each other. Ideally one would like to find a way to satisfy them all so the map can be computed in polynomial time and give, at worst, polynomially small probability to get a solution if the problem is soluble. One approach is to arrange for constructive interference in solutions while nonsolutions receive random contributions to their amplitude. While such random contributions are not as effective as a complete destructive interference, they are easier to construct and form the basis for a recent factoring algorithm [42] as well as the method presented here.

One possible mapping is based on analogy with backtracking search methods. Instead of examining just one path through the lattice of sets at a time, a superposition of states allows for considering all paths simultaneously. As with the division of search methods into a general strategy (e.g., backtrack) and problem specific choices, the quantum mapping described below has a general matrix that corresponds to exploring all possible changes to the partial sets, and a separate, particularly simple, matrix that incorporates information on the problem specific constraints. More complex maps are certainly possible, but this simple decomposition is easier to design and describe. Moreover, it suggests the possibility of implementing a special purpose quantum device to perform the general mapping, using the constraints of a specific problem only to adjust phases as described below.

## 3.3. Motivation

To motivate the mapping described below, we consider an idealized version of it to describe the general intuition of why paths through the lattice tend to interfere destructively for nonsolution states, provided the constraints are small.

The idealized map simply maps each set in the lattice equally to its supersets at the next level, while introducing random phases for sets found to be nogood. Note that for NP search problems, testing whether a particular set is nogood can be determined rapidly. For this discussion we are concerned with the relative amplitude in solutions and nogoods so we ignore the overall normalization. Thus for instance, with $N = 6$, the state $|\{1,2\}\rangle$ will map to an unnormalized superposition of its four supersets of size 3, namely the state $|\{1,2,3\}\rangle + \ldots + |\{1,2,6\}\rangle$.

With this mapping, a good at level $i$ will receive equal contribution from each of its $i$ subsets at the prior level. Starting with amplitude of 1 at level 0 then gives an amplitude of $i!$ for goods at level $i$. In particular, $L!$ for solutions.

How does this compare with contribution to nogoods, on average? This will depend on how many of the subsets are nogoods also. A simple case for comparison is when *all* sets in the lattice are nogood (starting with those at level $k$ given by the size of the constraints, e.g., $k = 2$ for problems with binary constraints). Let $r_i$ be the expected value of the magnitude of the amplitude for sets at level $i$. Each set at level $k$ will have $r_k = k!$ (and a zero phase) because all smaller subsets will be goods. A set $s$ at level $i > k$ will be a sum of $i$ contributions from (nogood) subsets, giving a total contribution of

$$\psi(s) = \sum_{m=1}^{i} \psi(s_m) e^{i\phi_m} \tag{5}$$

where the $s_m$ are the subsets of $s$ of size $i-1$ and the phases $\phi_m$ are randomly selected. The $\psi(s_m)$ have expected magnitude $r_{i-1}$ and some phase that can be combined with $\phi_m$ to give a new random phase $\theta_m$. Ignoring the variation in the magnitude of the amplitudes at each level this gives

$$r_i = r_{i-1} \left\langle \sum_{m=1}^{i} e^{i\theta_m} \right\rangle = r_{i-1} \sqrt{i} \tag{6}$$

because the sum of $i$ random phases is equivalent to an unbiased random walk [28] with $i$ unit steps which has expected net distance of $\sqrt{i}$. Thus we get $r_i = r_k \sqrt{i!/k!}$ or $r_i = \sqrt{i!k!}$ for $i > k$.

This crude argument gives a rough estimate of the relative probabilities for solutions compared to complete nogoods. Suppose there is only one solution. Then its relative probability is $L!^2$. The nogoods have relative probability $(N_L - 1)r_L^2 \sim N_L L! k!$ with $N_L$ given by Eq. 1. An interesting scaling regime is $L = n/b$ with fixed $b$, corresponding to a variety of well-studied constraint satisfaction problems. This gives

$$\ln\left(\frac{P_{soln}}{P_{nogood}}\right) = \ln\left(\frac{L!}{N_L k!}\right) \sim \frac{n}{b} \ln n + O(n) \tag{7}$$

This goes to infinity as problems get large so the enhancement of solutions is more than enough to compensate for their rareness among sets at the solution level.

The main limitation of this argument is assuming that all subsets of a nogood are also nogood. For many nogoods, this will not be the case, resulting in less opportunity for cancellation of phases. The worst situation in this respect is when most subsets are goods. This could be because the constraints are large, i.e., they don't rule out states until many items are included. Even with small constraints, this could happen occasionally due to a poor ordering choice for adding items to the sets, hence suggesting that a lattice restricted to assignments in a single order will be much less effective in canceling amplitude in nogoods. For the problems considered here, with small constraints, a large nogood cannot have too many good subsets because to be nogood means a small subset violates a (small) constraint and hence most subsets obtained by removing one element will still contain that bad subset giving a nogood. In fact, some numerical experiments (with the class of random problems described below) show that this mapping is very effective in canceling amplitude in the nogoods. Thus the assumptions made in this simplified argument seem to provide the correct intuitive description of the behavior.

Still the assumption of many nogood subsets underlying the above argument does suggest the extreme cancellation derived above will *least* apply when the problem has many large partial solutions. This gives a simple explanation for the difficulty encountered with the full map described below at the phase transition point: this transition is associated with problems with relatively many large partial solutions but few complete solutions. Hence we can expect relatively less cancellation of at least some nogoods at the solution level and a lower overall probability to find a solution.

This discussion suggests why a mapping of sets to supersets along with random phases introduced at each inconsistent set can greatly decrease the contribution to nogoods at the solution level. However, this mapping itself is not physically realizable because it is not unitary. For example, the mapping from level 1 to 2 with $N = 3$ takes the states $|\{1\}\rangle, |\{2\}\rangle, |\{3\}\rangle$ to $|\{1,2\}\rangle, |\{1,3\}\rangle, |\{2,3\}\rangle$ with the matrix

$$M = \begin{pmatrix} 1 & 1 & 0 \\ 1 & 0 & 1 \\ 0 & 1 & 1 \end{pmatrix} \qquad (8)$$

Here, the first column means the state $|\{1\}\rangle$ contributes equally to $|\{1,2\}\rangle$ and $|\{1,3\}\rangle$, its supersets, and gives no contribution to $|\{2,3\}\rangle$. We see immediately that the columns of this matrix are not orthogonal, though they can be easily normalized by dividing the entries by $\sqrt{2}$. More generally, this mapping takes each set at level $i$ to the $N-i$ sets with one more element. The corresponding matrix *M* has one column for each *i*–set and one row for each $(i+1)$-set. In each column there will be exactly $N-i$ 1's (corresponding to the supersets of the given *i*–set) and the remaining entries will be 0. Two columns will have at most a single nonzero value in common (and only when the two corresponding *i*–sets have all but one of their values in common: this is the only way they can share a superset in common). This means that as *N* gets large, the columns of this matrix are almost orthogonal (provided $i < N/2$, the case of interest here). This fact is used below to obtain a unitary matrix that is fairly close to *M*.

### 3.4. A Search Algorithm

The general idea of the mapping introduced here is to move as much amplitude as possible to supersets (just as in classical backtracking, increments to partial sets give supersets). This is combined with a problem specific adjustment of phases based on testing partial states for consistency (this corresponds to a diagonal matrix and thus is particularly simple in that it does not require any mixing of the amplitudes of different states). The specific methods used are described in this section.

### The Problem-Independent Mapping

To take advantage of the potential cancellation of amplitude in nogoods described above we need a unitary mapping whose behavior is similar to the ideal mapping to supersets. There are two general ways to adjust the ideal mapping of sets to supersets (mixtures of these two approaches are possible as well). First, we can keep some amplitude at the same level of the lattice instead of moving all the amplitude up to the next level. This allows using the ideal map described above (with suitable normalization) and so gives excellent discrimination between solutions and nonsolutions, but unfortunately not much amplitude reaches solution level. This is not surprising: the use of random phases cancel the amplitude in nogoods but this doesn't add anything to solutions (because solutions are not a superset of any nogood and hence cannot receive any amplitude from them). Hence at best, even when all nogoods cancel completely, the amplitude in solutions will be no more than their relative number among complete sets, i.e., very small. Thus the random phases prevent much amplitude moving to nogoods high in the lattice, but instead of contributing to solutions this amplitude simply remains at lower levels of the lattice. Hence we have no better chance than random selection of finding a solution (but, when a solution is not found, instead of getting a nogood at the solution level, we are now likely to get a smaller set in the lattice). Thus we must arrange for amplitude taken from nogoods to contribute instead to the goods. This requires the map to take amplitude to sets other than just supersets, at least to some extent.

The second way to fix the nonunitary ideal map is to move amplitude also to non-supersets. This can move all amplitude to the solution level. It allows some canceled amplitude from nogoods to go to goods, but also vice versa, resulting in less effective concentration into solutions. This can be done with a unitary matrix as close as possible to the nonunitary ideal map to supersets, and that also has a relatively simple form. The general question here is given *k* linearly independent vectors in *m* dimensional space, with $k \leq m$, find *k* orthonormal vectors in the space as close as possible to the *k* original ones. Restricting attention to the subspace defined by the original vectors, this can be obtained[4] using the singular value decomposition [22] (SVD) of the matrix *M* whose columns are the *k* given vectors. Specifically, this decomposition is $M = A^{\dagger} \Lambda B$, where $\Lambda$ is a diagonal matrix containing the singular values of *M* and both $A^{\dagger}$ and *B* have orthonormal columns. Then the matrix $U = A^{\dagger} B$ has orthonormal columns and is the closest set of orthogonal vectors according to the Frobenius matrix norm. For example,

---

[4]   I thank J. Gilbert for pointing out this technique, as a variant of the orthogonal Procrustes problem [22].

the mapping from level 1 to 2 with $N = 3$ given in Eq. 8 produces

$$U = \frac{1}{3} \begin{pmatrix} 2 & 2 & -1 \\ 2 & -1 & 2 \\ -1 & 2 & 2 \end{pmatrix} \qquad (9)$$

We should note that this construction fails if $k > m$ since an $m$–dimensional space cannot have more than $m$ orthogonal vectors. Hence we restrict consideration to mappings in the lattice at those levels $i$ where level $i + 1$ has at least as many sets as level $i$, i.e., $N_i \leq N_{i+1}$. Obtaining a solution requires mapping up to level $L$ so, from Eq. 1, this restricts consideration to problems where $L \leq \lceil N/2 \rceil$.

While this gives a set of orthonormal vectors close to the original map, one might be concerned about the requirement to compute the SVD of exponentially large matrices. Fortunately, however, the resulting matrices have a particularly simple structure in that the entries depend only on the overlap between the sets. Thus we have $U_{r\beta} = a_{|r \cap \beta|}$ ($r$ is an (i+1)-subset, $\beta$ is an i-subset). The overlap $|r \cap \beta|$ ranges from $i$ when $\beta \subset r$ to 0 when there is no overlap. Thus instead of exponentially many distinct values, there are only $i + 1$, a linear number. This can be exploited to give a simpler method for evaluating the entries of the matrix as follows.

We can get expressions for the $a$ values for a given $N$ and $i$ since the resulting column vectors are orthonormal. This gives

$$1 = \left(U^\dagger U\right)_{\alpha\alpha} = \sum_{k=0}^{i} n_k a_k^2 \qquad (10)$$

where

$$n_k = \binom{i}{k}\binom{N-i}{i+1-k} \qquad (11)$$

is the number of ways to pick $r$ with the specified overlap. For the off-diagonal terms, suppose $|\alpha \cap \beta| = p < i$ then

$$0 = \left(U^\dagger U\right)_{\alpha\beta} = \sum_{j,k=0}^{i} n_{jk}^{(p)} a_j a_k \qquad (12)$$

where

$$n_{jk}^{(p)} = \sum_x \binom{i-p}{k-x}\binom{p}{x}\binom{i-p}{j-x}\binom{N-2i+p}{i+1-j-k+x} \qquad (13)$$

is the number of sets $r$ with the required overlaps with $\alpha$ and $\beta$, i.e., $|r \cap \alpha| = k \leq i$ and $|r \cap \beta| = j \leq i$. In this sum, $x$ is the number of items the set $r$ has in common with both $\alpha$ and $\beta$. Together these give $i+1$ equations for the values of $a_0, \ldots, a_i$, which are readily solved numerically. Note in particular that the number of values and equations grows only linearly with the level in the lattice, even though the number of sets at each level grows exponentially. When necessary to distinguish the values at different levels in the lattice, we use $a_k^{(i)}$ to mean the value of $a_k$ for the mapping from level $i$ to $i + 1$.

There are multiple solutions for this system of quadratic equations, each representing a possible unitary mapping. But there is a unique one closest to the ideal mapping to supersets, as given by the SVD. It is this solution we use in the experiments, although an interesting open question is whether some other solution, in conjunction with various choices of phases, performs better.

A normalized version of the ideal map would have $a_i^{(i)} = \frac{1}{\sqrt{n_i}} = \frac{1}{\sqrt{N-i}}$ and all other values equal to zero. The actual values for $a_k^{(i)}$ are fairly close to this (confirming that the ideal map is close to orthogonal already), and alternate in sign. To illustrate their behavior, it is useful to consider the scaled values $b_k^{(i)} \equiv (-1)^k a_{i-k}^{(i)} \sqrt{n_{i-k}}$, with $n_{i-k}$ evaluated using Eq. 11. The behavior of these values for $N = 10$ is shown in Fig. 2. Note that $b_0^{(i)}$ is close to one, and decreases slightly as higher levels in the lattice (i.e., larger $i$ values) are considered: the ideal mapping is closer to orthogonal at low levels in the lattice.

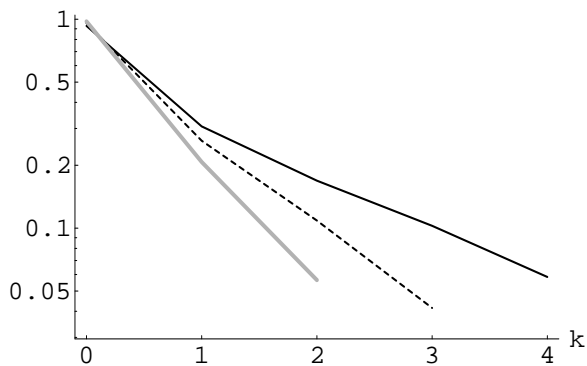

**Fig. 2.** Behavior of $b_k^{(i)}$ vs. $k$ on a log scale for $N = 10$. The three curves show the values for $i = 4$ (black), 3 (dashed) and 2 (gray).

Despite the simple values for the example of Eq. 9, the $a_k$ values in general do not appear to have a simple closed form expression. This is suggested by obtaining exact solutions to Eqs. 10 and 12 using a symbolic algebra program [49]. In most cases this gives complicated expressions involving nested roots. Since it is always possible such expressions could simplify, the $a_k$ values were also checked for being close to rational numbers and whether they are roots of single variable polynomials of low degree[5]. Neither simplification was found to apply.

Finally we should note that this mapping only describes how the sets at level $i$ are mapped to the next level. The full quantum system will also perform some mapping on the remaining sets in the lattice. By changing the map at each step, most of the other sets can simply be left unchanged, but there will need to be a map of the sets at level $i+1$ other than the identity mapping to be orthogonal to the map from level $i$. Any orthogonal set mapping partly back to level $i$ and partly remaining in sets at level $i+1$ will be suitable for this: in our application there is no amplitude at level $i+1$ when the map is used and hence it doesn't matter what mapping is used. However, the choice of this part of the overall mapping remains a degree of freedom that could perhaps be exploited to minimize errors introduced by external noise.

---

[5] Using the Mathematica function Rationalize and the package NumberTheory'Recognize'.

## Phases for Nogoods

In addition to the general mapping from one level to the next, there is the problem-specific aspect of the algorithm, namely the choice of phases for the nogood sets at each level. In the ideal case described above, random phases were given to each nogood, resulting in a great deal of cancellation for nogoods at the solution level. While this is a reasonable choice for the unitary mapping, other policies are possible as well. For example, one could simply invert the phase of each nogood[6] (i.e., multiply its amplitude by -1). This choice doesn't work well for the idealized map to supersets only but, as shown below, is helpful for the unitary map. It can be motivated by considering the coefficients shown in Fig. 2. Specifically, when a nogood is encountered for the first time on a path through the lattice, we would like to cancel phase to its supersets but at the same time enhance amplitude in other sets likely to lead to solutions. Because $a_{i-1}^{(i)}$ is negative, inverting the phase will tend to add to sets that differ by one element from the nogood. At least some of these will avoid violating the small constraint that produced this nogood set, and hence may contribute eventually to sets that do lead to solutions.

Moreover, one could use information on the sets at the next level to decide what to do with the phase: as currently described, the computation makes no use of testing the consistency of sets at the solution level itself, and hence is completely ineffective for problems where the test requires the complete set. Perhaps better would be to mark a state as nogood if it has no consistent extensions with one more item (this is simple to check since the number of extensions grows only linearly with problem size). Another possibility is for the phase to be adjusted based on how many constraints are violated, which could be particularly appropriate for partial constraint satisfaction problems [19] or other optimization searches.

## Summary

The search algorithm starts by evenly dividing amplitude among the goods at level 2 of the lattice (whose number is proportional to $N^2$). Then for each level from 2 to $L-1$, we adjust the phases of the states depending on whether they are good or nogood and then map to the next level. Thus if $\psi_\alpha^{(i)}$ represents the amplitude of the set $\alpha$ at level $i$, we have

$$\psi_r^{(i+1)} = \sum_\alpha U_{r\alpha} \rho_\alpha \psi_\alpha^{(i)} = \sum_k a_k^{(i)} \sum_{|r \cap \alpha|=k} \rho_\alpha \psi_\alpha^{(i)} \qquad (14)$$

where $\rho_\alpha$ is the phase assigned to the set $\alpha$ after testing whether it is nogood, and the final inner sum is over all sets $\alpha$ that have $k$ items in common with $r$. That is, $\rho_\alpha = 1$ when $\alpha$ is a good set. For nogoods, $\rho_\alpha = -1$ when using the phase inversion method, and $\rho_\alpha = e^{i\theta}$ with $\theta$ uniformly selected from $[0, 2\pi)$ when using the random phase method. Finally we measure the state, obtaining a complete set. This set will be a solution with probability $p_{soln} = \sum_s \psi_s^{(L)}$, with the sum over solution sets, depending on the particular problem and method for selecting the phases.

---

[6] I thank J. Lamping for suggesting this.

What computational resources are required for this algorithm? The storage requirements are quite modest: $N$ bits can produce a superposition of $2^N$ states, enough to represent all the possible sets in the lattice structure. Since each trial of this algorithm gives a solution only with probability $p_{soln}$, on average it will need to be repeated $1/p_{soln}$ times to find a solution. The cost of each trial consists of the time required to construct the initial superposition and then evaluate the mapping on each step from the level 2 to the solution level $L$, a total of $L - 2 < N/2$ mappings. Because the initial state consists of sets of size 2, there are only a polynomial number of them (i.e., $O(N^2)$) and hence the cost to construct the initial superposition will be relatively modest. The mapping from one level to the next will need to be produced by a series of more elementary operations, namely unitary matrices with only a fixed number of nonzero entries. Determining the required number of such operations remains an open question, though the particularly simple structure of the matrices should not require involved computations and should also be able to exploit special purpose hardware. At any rate, this mapping is independent of the structure of the problem and its cost does not affect the relative costs of different problem structures. Finally, determining the phases to use for the nogood sets involves testing the sets against the constraints, a relatively rapid operation for NP search problems. Thus to examine how the cost of this search algorithm depends on problem structure, the key quantity is the behavior of $p_{soln}$.

### 3.5. An Example

To illustrate the algorithm's operation and behavior, consider the small case of $N = 3$ with the map starting from level 0 and going up to level 2. Suppose that $\{3\}$ and its supersets are the only nogoods. For the purposes of illustration, we begin with all amplitude in the empty set, i.e., with the state $|\emptyset\rangle$, rather than in the goods at level 2. The map from level 0 to 1 gives equal amplitude to all singleton sets, producing $\frac{1}{\sqrt{3}}(|\{1\}\rangle + |\{2\}\rangle + |\{3\}\rangle)$. We then introduce a phase for the nogood set, giving $\frac{1}{\sqrt{3}}(|\{1\}\rangle + |\{2\}\rangle + e^{i\theta}|\{3\}\rangle)$. Finally we use Eq. 9 to map this to the sets at level 2, giving the final state

$$\frac{1}{3\sqrt{3}}\left(\left(4 - e^{i\theta}\right)|\{1,2\}\rangle + \left(1 + 2e^{i\theta}\right)|\{1,3\}\rangle + \left(1 + 2e^{i\theta}\right)|\{2,3\}\rangle\right) \quad (15)$$

At this level, only set $\{1,2\}$ is good, i.e., a solution. Note that the algorithm does not make any use of testing the states at the solution level for consistency.

The probability to obtain a solution when the final measurement is made is determined by the amplitude of the solution set, so in this case

$$p_{soln} = \left|\frac{1}{3\sqrt{3}}\left(4 - e^{i\theta}\right)\right|^2 = \frac{1}{27}(17 - 8\cos\theta) \quad (16)$$

From this we can see the effect of different methods for selecting the phase for nogoods. If the phase is selected randomly, $p_{soln} = \frac{17}{27} = 0.63$ because the average value of $\cos\theta$ is zero. Inverting the phase of the nogood, i.e., using $\theta = \pi$, gives $p_{soln} = \frac{25}{27} = 0.93$. These probabilities compare with the 1/3 chance of selecting a solution by random choice. In this case, the optimal choice of phase is the same as that obtained by simple inversion. However this is not true in general: picking phases optimally will require knowledge about the solutions and hence is not a feasible mapping. Note also that even the optimal choice of phase doesn't guarantee a solution is found.

## 4. Average Behavior of the Algorithm

For classical simulation of this algorithm we explicitly compute the amplitude of all sets in the lattice up to the solution level resulting in an exponential slowdown compared to the quantum behavior, and even with respect to more efficient classical search methods that halt after finding the first solution. Unfortunately, the requirements to evaluate all sets in the lattice and the mappings between them severely limit the feasible size of these classical simulations. Moreover, to simulate the expected behavior of the random phase method, we must repeat the search a number of times on each problem (10 tries in the experiments reported here) to estimate the average behavior with respect to the selection of the random phases. This further limits the feasible problem size. By contrast, the phase inversion method determines the probability to find a solution with a single trial and hence allows exploration of somewhat larger problems.

As a simple check on the numerical errors of the calculation, we recorded the total normalization in all sets at the solution level. With double precision calculations on a Sun Sparc10, for the experiments reported here typically the norm was 1 to within a few times $10^{-11}$. As an indication of the execution time with unoptimized C++ code for the experiments reported below, a single trial for a problem with $N = 14$ and 16, with $L = N/2$, required about 70 and 1000 seconds, respectively. This uses a direct evaluation of the map from one level to the next as given by Eq. 14. A substantial reduction in compute time is possible by exploiting the simple structure of this matrix to give a recursive evaluation[7]. Some additional improvement is possible by better coding and exploiting the fact that all amplitudes are real when using the method that inverts phases of nogoods.

### 4.1. A Class of Search Problems

To examine the typical behavior of this quantum search algorithm with respect to problem structure, we need a suitable class of problems. This is particularly important for average case analyses since one could inadvertently select a class of search problems dominated by easy cases. Fortunately the observed concentration of hard cases near phase transitions provides a method to generate hard test cases.

---

[7] I thank S. Vavasis for suggesting this improvement in the classical simulation of the algorithm.

The phase transition behavior has been seen in a variety of search problem classes. Here we select a particularly simple class of problems by supposing the constraints specify nogoods randomly at level 2 in the lattice. This corresponds to binary constraint satisfaction problems [39, 45], but ignores the detailed structure of the nogoods imposed by the requirement that variables have a unique assignment. By ignoring this additional structure, we are able to test a wider range of the number of specified nogoods for the problems than would be the case by considering only constraint satisfaction problems. This lack of additional structure is also likely to make the asymptotic behavior more readily apparent at the small problem sizes that are feasible with a classical simulation.

Furthermore, since the quantum search algorithm is appropriate only for soluble problems, we restrict attention to random problems with a solution. These could be obtained by randomly generating problems and rejecting any that have no solution (as determined using a complete classical search method). However, for overconstrained problems the soluble ones become quite rare and difficult to find by this method. Instead, we generate problems with a prespecified solution. That is, when randomly selecting nogoods to add to a problem, we do not pick any nogoods that are subsets of a prespecified solution set. This always produces problems with at least one solution. Although these problems tend to be a bit easier than randomly selected soluble problems, they nevertheless exhibit the same concentration of hard problems and at about the same location as general random problems [7, 48].

For our class of problems, this behavior is illustrated in Fig. 3. Specifically, this shows the cost to solve the problem with a simple chronological backtrack search. The cost is given in terms of the number of decision points considered until a solution is found. The minimum cost, for a search that proceeds directly to a solution with no backtrack is $L + 1$. The parameter distinguishing underconstrained from overconstrained problems is the ratio $\beta$ of the number of nogoods $m$ at level 2 given by the constraints to the number of items $N$.

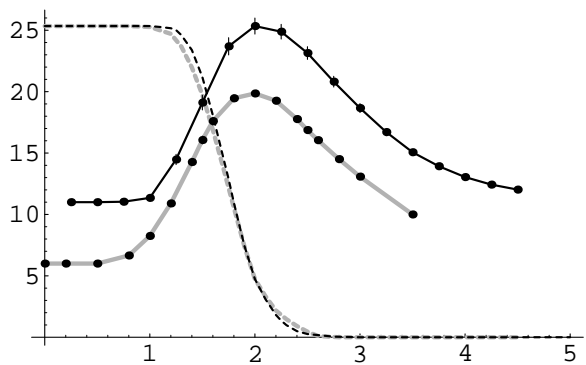

**Fig. 3.** The solid curves show the classical backtrack search cost for randomly generated problems with a prespecified solution as a function of $\beta = m/N$ for $N = 10$ (gray) and 20 (black) and $L = N/2$. Here $m$ is the number of nogoods selected at level 2 of the search lattice. The cost is the average number of backtrack steps, starting from the empty set, required to find the first solution to the problem, averaged over 1000 problems. The error bars indicate the standard deviation of this estimate of the average value, and in most cases are smaller than the size of the plotted points. For comparison, the dashed curves show the probability for having a solution in randomly generated problems with the specified $\beta$ value, ranging from 1 at the left to 0 at the right.

Even for these relatively small problems, a peak in the average search cost is evident. Moreover, this peak is near the critically constrained region where random problems change from mostly soluble to mostly insoluble. A simple, but approximate, theoretical value for the location of the transition region is given by the point where the expected number of solutions is equal to one [45, 48]. Applying this to the class of problems considered here is straightforward. Specifically, there are $N_L$ complete sets for the problem, as given by Eq. 1. For a particular set $s$ of size $L$, it will be good, i.e., a solution, only if none of the nogoods selected for the problem are a subset of $s$. Hence the probability it will be a solution is given by

$$\rho_L = \frac{\binom{\binom{N}{2}-\binom{L}{2}}{m}}{\binom{\binom{N}{2}}{m}} \quad (17)$$

because there are $\binom{N}{2}$ sets of size 2 from which to choose the $m$ nogoods specified directly by the constraints. The average number of solutions is then just $N_{soln} = N_L \rho_L$. If we set $m = \beta N$ and $L = N/b$, for large $N$ this becomes

$$\ln N_{soln} \sim N \left( h\left(\frac{1}{b}\right) + \beta \ln \left(1 - \frac{1}{b^2}\right) \right) \quad (18)$$

where $h(x) \equiv -x \ln x - (1-x) \ln (1-x)$. The predicted transition point[8] is then given by

$$\beta_{crit} = \frac{h(1/b)}{\ln(1 - 1/b^2)} \quad (19)$$

which is $\beta_{crit} = 2.41$ for the case considered here (i.e., $b = 2$). This closely matches the location of the peak in the search cost for problems with prespecified solution, as shown in Fig. 3, but is about 20% larger than the location of the step in solubility[9]. Furthermore, the theory predicts there is a regime of polynomial average cost for sufficiently few constraints [26]. This is determined by the condition that the expected number of goods at each level in the lattice is monotonically increasing. Repeating the above argument for smaller levels in the lattice, we find that this condition holds up to

$$\beta_{poly} = \frac{1 - b^2}{2b} \ln \frac{1}{b-1} \quad (20)$$

which is $\beta_{poly} = 0$ for $b = 2$.

While these estimates are only approximate, they do indicate that the class of random soluble problems defined here behaves qualitatively and quantitatively the same with respect to the transition behavior as a variety of other, perhaps more realistic, problem classes. This close correspondence with the theory (derived for the limit of large problems), suggests that we are observing the correct transition behavior even with these relatively small problems. Hence it is a reasonable basis for evaluating the behavior of the quantum search algorithm. Moreover the above approximate theoretical argument suggests that the average cost of general classical search methods scales exponentially with the size of the problem over the full range of $\beta > 0$. Thus this provides a good test case for the average behavior of the quantum algorithm. As a final observation, it is important to obtain a sufficient number of samples especially near the transition region. This is because there is considerable variation in problems near the transition, specifically a highly skewed distribution in the number of solutions. In this region, most problems have few solutions but a few have extremely many: enough in fact to give a substantial contribution to the average number of solutions even though such problems are quite rare.

---

[8] This differs slightly from the results for problems with more specified structure on the nogoods, such as explicitly removing the necessary nogoods from consideration [45, 48].

[9] This is a particularly large error for this theory: it does better for problems with larger constraints or more allowed values per variable.

## 4.2. Phase Transition

To see how problem structure affects this search algorithm, we evaluate $p_{soln}$, the probability to find a solution for problems with different structures, ranging from underconstrained to overconstrained. Low values for this probability indicate relatively harder problems. The expected number of repetitions of the search required to find a solution is then given by $1/p_{soln}$. The results are shown in Figs. 4 and 5 for different ways of introducing phases for nogood sets. We see the general easy-hard-easy pattern in both cases.

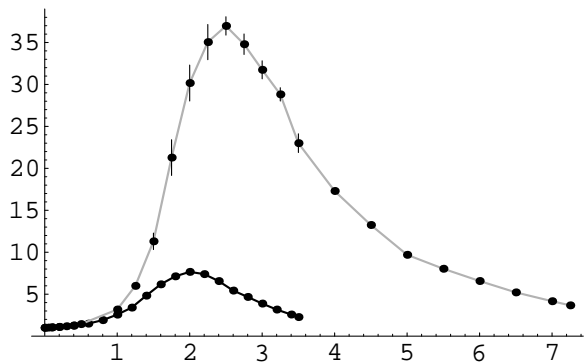

**Fig. 4.** Expected number of trials to find a solution vs. $\beta$ for random problems with prespecified solution with binary constraints, using random phases for nogoods. The solid curve is for $N = 10$, with 100 samples per point. The gray curve is for $N = 20$ with 10 samples per point (except that additional samples were used around the peak). The error bars indicate the standard error in the estimate of $\langle 1/p_{soln} \rangle$.

Another common feature of phase transitions is an increased variance around the transition region. For this method we see a peak in the variance as well, shown in Fig. 6. Furthermore, there is no indication of the rare hard cases in the easy region of underconstrained problems [21, 26, 44]. While this could be due to the small cases examined here, it is more likely that this method, by considering simultaneously all ways for constructing sets moving up the lattice, cannot get stuck with a poor initial choice leading to thrashing (i.e., in classical backtrack methods, when the initial choices happen to leave many large partial solutions but most other choices do not). Hence this provides another example of a distinct class of algorithms, as with more sophisticated backtracking [1], that appears to avoid hard problems in the underconstrained region.

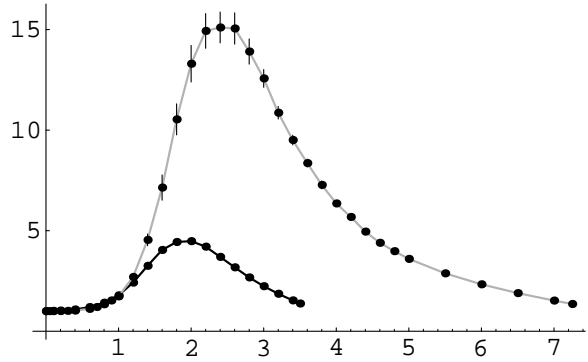

**Fig. 5.** Expected number of trials to find a solution vs. $\beta$ for random problems with prespecified solution with binary constraints, using inverted phases for nogoods. The solid curve is for $N = 10$, with 1000 samples per point. The gray curve is for $N = 20$ with 100 samples per point (except that additional samples were used around the peak). The error bars indicate the standard error in the estimate of $\langle 1/p_{soln} \rangle$.

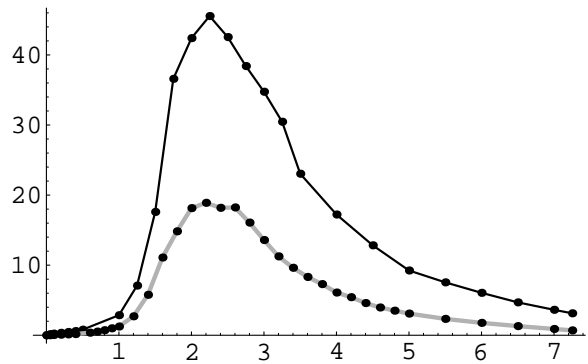

**Fig. 6.** Standard deviation in the number of trials to find a solution for $N = 20$ as a function of $\beta$. The black curve is for random phases assigned to nogoods, and the gray one for inverting phases.

### 4.3. Scaling

An important question in the behavior of this search method is how its average performance scales with problem size. To examine this question, we consider the scaling with fixed $\beta$. This is shown in Figs. 7 and 8 for algorithms using random and inverted phases for nogoods, respectively. We show the results on both a log plot (where straight lines correspond to exponential scaling) and a log-log plot (where straight lines correspond to power-law or polynomial scaling).

It is difficult to make definite conclusions from these results for two reasons. First, the variation in behavior of different problems gives a statistical uncertainty to the estimates of the average values, particularly for the larger sizes where fewer samples are available. The standard errors in the estimates of the averages are indicated by the error bars in the figures (though in most cases, the errors are smaller than the size of the plotted points). Second, the scaling behavior could change as larger cases are considered. With these caveats in mind, the figures suggest that $p_{soln}$ remains nearly constant for underconstrained problems, even though the fraction of complete sets that are solutions is decreasing exponentially. This behavior is also seen in the overlap of the curves for small $\beta$ in Figs. 4 and 5. For problems with more constraints, $p_{soln}$ appears to decrease polynomially with the size of the problem. An interesting observation in comparing the two phase choices is that the scaling is qualitatively similar. This suggests the detailed values of the phase choices are not critical to the scaling behavior, and in particular high precision evaluation of the phases is not required. Finally we should note that this illustration of the average scaling leaves open the behavior for the worst case instances.

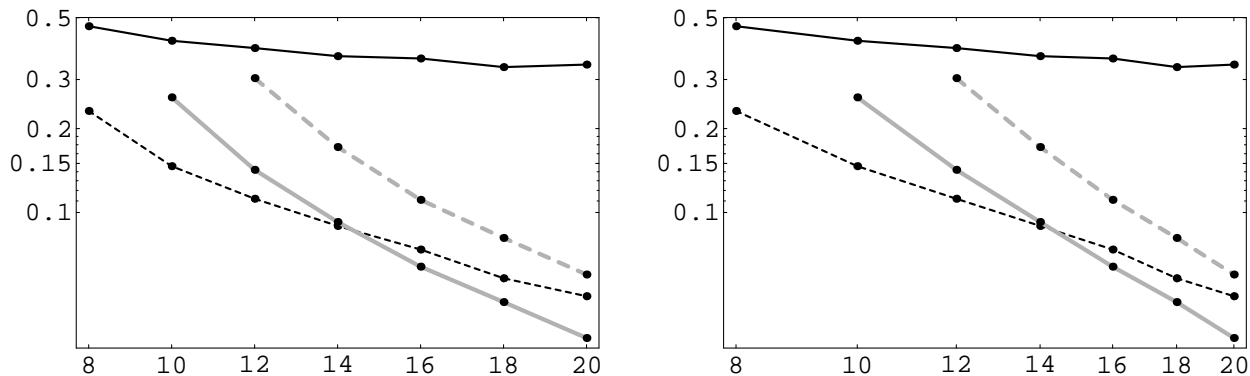

**Fig. 7.** Scaling of the probability to find a solution using the random phase method, for $\beta$ of 1 (solid), 2 (dashed), 3 (gray) and 4 (dashed gray). This is shown on log and log-log scales (left and right plots, respectively).

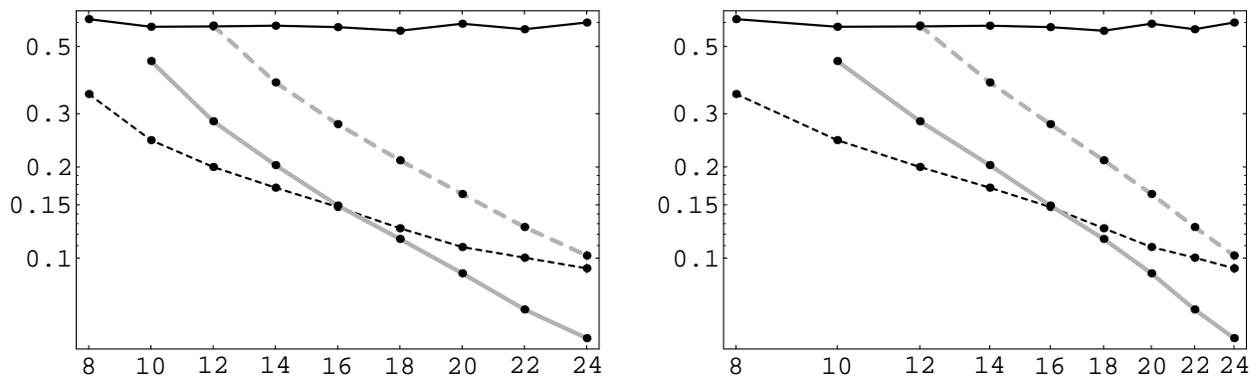

**Fig. 8.** Scaling of the probability to find a solution using the phase inversion method, for $\beta$ of 1 (solid), 2 (dashed), 3 (gray) and 4 (dashed gray). This is shown on log and log-log scales (left and right plots, respectively).

Another scaling comparison is to see how much this algorithm enhances the probability to find a solution beyond the simple quantum algorithm of evaluating all the complete sets and then making a measurement. As shown in Fig. 9, this appears to give an exponential improvement in the concentration of amplitude into solutions.

### 4.4. Random 3SAT

These experiments leave open the question of how additional problem structure might affect the scaling behaviors. While the universality of the phase transition behavior suggests that the average behavior of this algorithm will also be the same for a wide range of problems, it is useful to check this empirically. To this end the algorithm was applied to the satisfiability (SAT) problem. This constraint satisfaction problem consists of a propositional formula with $n$ variables and the requirement to find an assignment (true or false) to each variable that makes the formula true. Thus there are $b = 2$ assignments for each variable and $N = 2^n$ possible assignments. We consider the well-studied NP-complete 3SAT problem where the formula is a conjunction of $c$ clauses, each of which is a disjunction of 3 (possibly negated) variables.

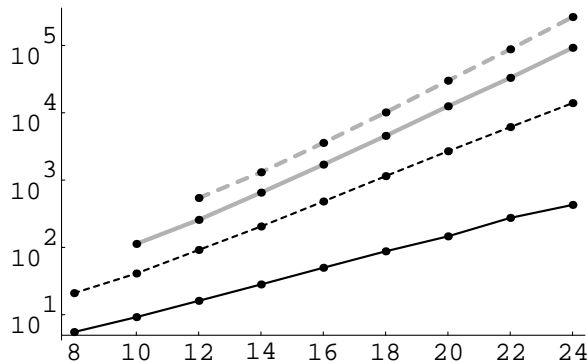

**Fig. 9.** Scaling of the ratio of the probability to find a solution using the quantum algorithm to the probability to find a solution by random selection at the solution level, using the phase inversion method, for $\beta$ of 1 (solid), 2 (dashed), 3 (gray) and 4 (dashed gray). The curves are close to linear on this log scale indicating exponential improvement over the direct selection from among complete sets, with a higher enhancement for problems with more constraints.

The SAT problem is readily represented in terms of nogoods in the lattice of sets [48]. As described in Sec. 2.2, there will be $n$ necessary nogoods, each of size 2. In addition, each distinct clause in the proposition gives a single nogood of size 3. This case is thus of additional interest in having specified nogoods of two sizes. For evaluating the quantum algorithm, we start at level 3 in the lattice. Thus the smallest case for which the phase choices will matter is for $n = 5$.

We generate random problems with a given number of clauses by selecting that number of different nogoods of size 3 from among those not already excluded by the necessary nogoods[10]. For random 3SAT, the hard problems are concentrated near the transition [36] at $c = 4.2n$. We investigated a range of values for $c/n$: 2, 4, 6 and 8. Finally, from among these randomly generated problems, we use only those that do in fact have a solution[11]. Using randomly selected soluble problems, rather than a prespecified solution results in somewhat harder problems. Like other studies that need to examine many goods and nogoods in the lattice [40], these results are restricted to much smaller problems than in most studies of random SAT. Consequently, the transition region is rather spread out. Furthermore, the additional structure of the necessary nogoods and the larger size of the constraints, compared with the previous set of problems, makes it more likely that larger problems will be required to see the asymptotic scaling behavior. However, at least some asymptotic behaviors have been observed [10] to persist quite accurately even for problems as small as $n = 3$, so some indication of the scaling behavior is not out of the question for the small problems considered here.

The resulting scaling of the probability to find a solution is shown in Fig. 10 using the phase inversion method. More limited experiments with the random phase method showed the same behavior as seen with the previous set of problems: somewhat worse performance but similar scaling behavior. The results here are less clear cut than those of Fig. 8. For $c/n = 2$ the results are consistent with either polynomial or exponential scaling. For problems with more constraints, exponential scaling is a somewhat better fit.

---

[10] This differs slightly from other studies of random 3SAT in not allowing duplicate clauses in the propositional formula.
[11] For the ratios of $c/n$ and small problems examined here, there are many soluble instances. Thus, there is no need to rely on a prespecified solution in order to efficiently find soluble test problems.

More definite results are obtained for the improvement over random selection. Specifically, Fig. 11 shows an exponential improvement for both the phase inversion and random phase methods, corresponding to the behavior for random problems in Fig. 9. Similar improvement is seen for other values of $c/n$ as well: as in Fig. 9 the more highly constrained problems give larger improvements. A more stringent comparison is with random selection from among valid assignments (i.e., each variable given a single value) rather than from among general sets of assignments. This is also shown in Fig. 11, appearing to grow exponentially as well. This is particularly significant because the quantum algorithm makes no use of the possibility of explicitly removing from consideration those sets that give multiple assignments to the same variable. We conclude from these results that the additional structure of necessary nogoods and constraints of different sizes is qualitatively similar to that for unstructured random problems but a detailed comparison of the scaling behaviors requires examining larger problem sizes.

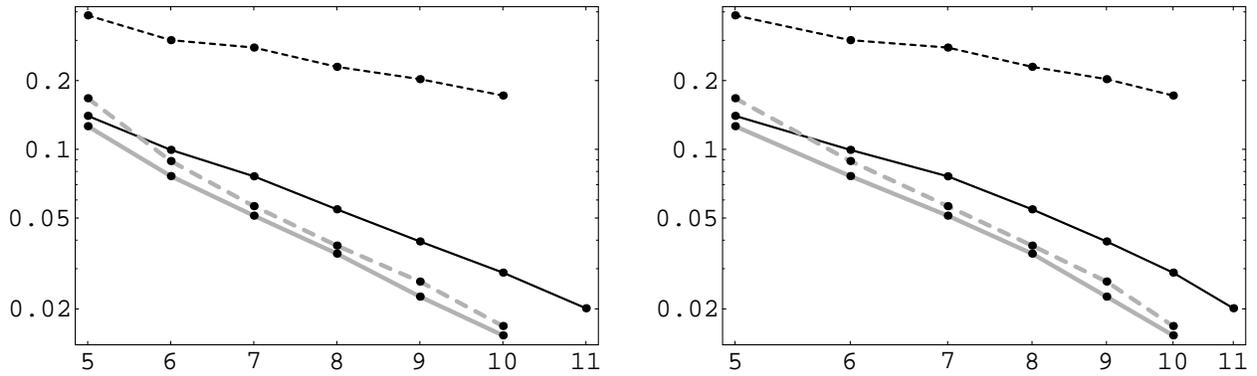

**Fig. 10.** Scaling of the probability to find a solution, using the phase inversion method, as a function of the number of variables for random 3SAT problems. The curves correspond to different clause to variable ratios: 2 (dashed), 4 (solid), 6 (gray) and 8 (gray, dashed). This is shown on log and log-log scales (left and right plots, respectively).

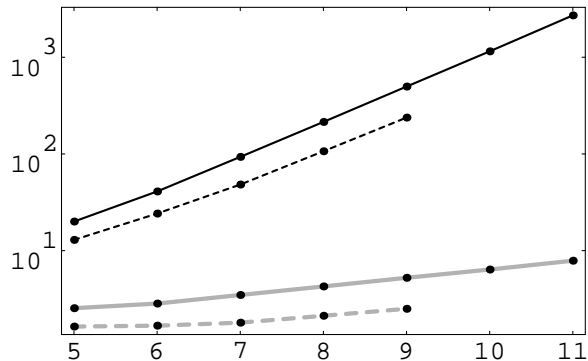

**Fig. 11.** Scaling of the ratio of the probability to find a solution using the quantum algorithm to the probability to find a solution by random selection at the solution level as a function of the number of variables for random 3SAT problems with clause to variable ratio equal to 4. The solid and dashed curves correspond to using the phase inversion and random phase methods, respectively. The black curves compare to random selection among complete sets, while the gray compare to selection only from among complete assignments. The curves are close to linear on this log scale indicating exponential improvement over the direct selection from among complete sets.

# 5. Discussion

In summary, we have introduced a quantum search algorithm and evaluated its average behavior on a range of small search problems. It appears to increase the amplitude into solution states exponentially compared to evaluating and measuring a quantum superposition of potential solutions directly. Moreover, this method exhibits the same transition behavior, with its associated concentration of hard problems, as seen with many classical search methods. It thus extends the range of methods to which this phenomenon applies. More importantly, this indicates the algorithm is effectively exploiting the same structure of search problems as, say, classical backtrack methods, to prune unproductive search directions. It is thus a major improvement over the simple applications of quantum computing to search problems that behave essentially the same as classical generate-and-test, a method that completely ignores the possibility of pruning and hence doesn't exhibit the phase transition.

The transition behavior is readily understood because problems near the transition point have many large partial goods that do not lead to solutions [48]. Thus there will be a relatively high proportion of paths through the lattice that appear good for quite a while but eventually give deadends. A choice of phases based on detecting nogoods will not be able to work on these paths until near the solution level and hence give less chance to cancel out or move amplitude to those paths that do in fact lead to solutions. Hence problems with many large partial goods are likely to prove relatively difficult for any quantum algorithms that operate by distinguishing goods from nogoods of various sizes.

There remain a large number of open questions. In the algorithm, the division between a problem–independent mapping through the lattice and a simple problem-specific adjustment to phases allows for a range of policies for selecting the phases. It would be useful to understand the effect of different policies in the hope of improving the concentration of amplitude into solutions. For example, the use of phases has two distinct jobs: first, to keep amplitude moving up along good sets rather than diffusing out to nogoods, and second, when a deadend is reached (i.e., a good set that has no good supersets) to send the amplitude at this deadend to a promising region of the search space, possibly very far from where the deadend occurred. These goals, of keeping amplitude concentrated on the one hand and sending it away on the other, are to some extent contradictory. Thus it may prove worthwhile to consider different phase choice policies for these two situations. Furthermore, the mapping through the lattice is motivated by classical backtrack methods, which move from sets to supersets in the lattice. It may also prove fruitful to consider another type of mapping based on local repair methods moving among neighbors of complete sets. In this case, sets are evaluated based on the number of constraints they violate so an appropriate phase selection policy could depend on this number, rather than just whether the set is inconsistent or not. These possibilities may also suggest new probabilistic classical algorithms that might be competitive with existing heuristic search methods (which are exponentially slow).

As a new example of a search method exhibiting the transition behavior, this work raises the same issues as prior studies of this phenomenon. For instance, to what extent does this behavior apply to more realistic classes of problems? For instance including the structure due to prohibiting multiple assignments in constraint satisfaction problems [39, 45] or clustering inherent in situations involving localized interactions [24]. This will be difficult to check empirically due to the limitation to small problems that are feasible for a classical simulation of this algorithm. However the observation that this behavior persists for many classes of problems with other search methods suggests it will be widely applicable. It is also of interest to see if other phase transition phenomena appear in these quantum search algorithms, such as observed in optimization searches [7, 38, 50]. There may also be transitions unique to quantum algorithms, for example in the required coherence time or sensitivity to environmental noise.

For the specific instances of the algorithm presented here, there are also some remaining issues. An important one is the cost of the mapping from one level to the next in terms of more basic operations that might potentially be realized in hardware, although the simple structure of the matrices involved suggest this should not be too costly. The scaling behavior of the algorithm for larger cases is also of interest, which can perhaps be approached by examining the asymptotic nature of the matrix coefficients of Eqs. 10 and 12.

An important practical question is the physical implementation of quantum computers in general [2, 43, 9], and the requirements imposed by the algorithm described here. Any implementation of a quantum computer will need to deal with two important difficulties [32]. First, there will be defects in the construction of the device. Thus even if an ideal design exactly produces the desired mapping, occasional manufacturing defects and environmental noise will introduce errors. We thus need to understand the sensitivity of the algorithm's behavior to errors in the mappings. Here the main difficulty is likely to be in the problem-independent mapping from one level of the lattice to the next, since the choice of phases in the problem-specific part doesn't require high precision. In this context we should note that standard error correction methods cannot be used with quantum computers in light of the requirement that all operations are reversible. We also need to address the extent to which such errors can be minimized in the first place, thus placing less severe requirements on the algorithm. Particularly relevant in this respect is the possibility of drastically reducing defects in manufactured devices by atomically precise control of the hardware [14, 15, 37]. This could substantially extend the range of ideal quantum algorithms that will be possible to implement.

The second major difficulty with constructing quantum computers is maintaining coherence of the superposition of states long enough to complete the computation. Environmental noise gradually couples to the state of the device, reducing the coherence and eventually limiting the time over which a coherent superposition can perform useful computations [47, 8]. In effect, the coupling to the environment can be viewed as performing a measurement on the quantum system, destroying the superposition of states. This problem is particularly severe for proposed universal quantum computers that need to maintain superpositions for arbitrarily long times. Decoherence is, in principle, less of an issue for algorithms whose execution time can be specified in advance. This is the case for the method presented here since it involves a fixed number of maps to move up the lattice structure, and could be implemented as a special purpose search device (for problems of a given size) rather than as a program running on a universal computer. Thus a given achievable coherence time would translate into a limit on feasible problem size. To the extent that this limit can be made larger than feasible for alternative classical search methods, the quantum search could be useful.

The open question of greatest theoretical interest is whether this algorithm or simple variants of it can concentrate amplitude into solutions sufficiently to give a polynomial, rather than exponential, decrease in the probability to find a solution of *any* NP search problem with small constraints. This is especially interesting since this class of problems includes many well-studied NP-complete problems such as graph coloring and propositional satisfiability. Even if this is not so in the worst case, it may be so on average for some classes of otherwise difficult real-world problems. While it is by no means clear to what extent quantum coherence provides more powerful computational behavior than classical machines, a recent proposal for rapid factoring [42] is an encouraging indication of its capabilities.

A more subtle question along these lines is how the average scaling behaves away from the transition region of hard problems. In particular, can such quantum algorithms expand the range of the polynomially scaling problems seen for highly underconstrained or overconstrained instances? If so, this would provide a class of problems of intermediate difficulty for which the quantum search is exponentially faster than classical methods, on average. This highlights the importance of broadening theoretical discussions of quantum algorithms to include typical or average behaviors in addition to worst case analyses. More generally, are there any differences in the phase transition behaviors or their location compared with the usual classical methods? These questions, involving the precise location of transition points, are not currently well understood even for standard classical search algorithms. Thus a comparison with the behavior of this quantum algorithm may help shed light on the nature of the various phase transitions that seem to be associated with the intrinsic structure of the search problems rather than with specific search algorithms.

### Acknowledgments

I thank John Gilbert, John Lamping and Steve Vavasis for their suggestions and comments on this work. I have also benefited from discussions with Peter Cheeseman, Scott Clearwater, Bernardo Huberman, Don Kimber, Colin Williams, Andrew Yao and Michael Youssefmir.